\documentclass[12pt,preprint]{aastex63}

\usepackage{graphicx}
\usepackage{amsmath, amssymb, amsthm}
\usepackage{multirow}
\usepackage{upgreek}
\usepackage{xcolor}

\def\lapp{\ifmmode\stackrel{<}{_{\sim}}\else$\stackrel{<}{_{\sim}}$\fi}
\def\gapp{\ifmmode\stackrel{>}{_{\sim}}\else$\stackrel{>}{_{\sim}}$\fi}

\newcommand{\rthree}{FRB~180916.J0158+65}
\newcommand{\repeater}{FRB~121102}  
\newcommand{\source}{\rthree}
\newcommand{\src}{\source}

\newcommand{\cxo}{\textit{Chandra}}
\newcommand{\cxofull}{\textit{Chandra X-ray Observatory}}
\newcommand{\chandra}{\textit{Chandra}}

\newcommand{\swift}{\textit{Swift}}
\newcommand{\fermi}{\textit{Fermi}}
\newcommand{\xmm}{\textit{XMM-Newton}}

\newcommand{\degrees}{^{\circ}}

\newcommand{\dmunits}{\,pc\,cm$^{-3}$}
\newcommand{\nhunits}{\,cm$^{-2}$}
\newcommand{\nh}{\ensuremath{N_\mathrm{H}}}
\newcommand{\fluxcgs}{erg~s$^{-1}$~cm$^{-2}$}
\newcommand{\flucgs}{erg~cm$^{-2}$}
\newcommand{\lumcgs}{erg~s$^{-1}$}

\newcommand{\ps}{}

\begin{document}

\title{Simultaneous X-ray and Radio Observations of the Repeating Fast Radio Burst \src}

\correspondingauthor{P. Scholz}
\email{paul.scholz@dunlap.utoronto.ca}

\author[0000-0002-7374-7119]{P. Scholz}
  \affiliation{Dunlap Institute for Astronomy \& Astrophysics, University of Toronto, 50 St.~George Street, Toronto, ON M5S 3H4, Canada}
\author[0000-0001-6422-8125]{A. Cook}
  \affiliation{Dunlap Institute for Astronomy \& Astrophysics, University of Toronto, 50 St.~George Street, Toronto, ON M5S 3H4, Canada}
  \affiliation{David A.~Dunlap Institute Department of Astronomy \& Astrophysics, University of Toronto, 50 St.~George Street, Toronto, ON M5S 3H4, Canada}
\author{M. Cruces}
  \affiliation{Max Planck Institut f\"{u}r Radioastronomie, Auf dem H\"{u}gel 69, D-53121, Bonn, Germany}
\author[0000-0003-2317-1446]{J.~W.~T. Hessels}
  \affiliation{Anton Pannekoek Institute for Astronomy, University of Amsterdam, Science Park 904, 1098 XH, Amsterdam, The Netherlands}
  \affiliation{ASTRON, Netherlands Institute for Radio Astronomy, Oude Hoogeveensedijk 4, 7991 PD Dwingeloo, The Netherlands}
\author[0000-0001-9345-0307]{V.~M. Kaspi}
  \affiliation{Department of Physics, McGill University, 3600 rue University, Montr\'eal, QC H3A 2T8, Canada}
  \affiliation{McGill Space Institute, McGill University, 3550 rue University, Montr\'eal, QC H3A 2A7, Canada}
\author[0000-0002-4694-4221]{W.~A. Majid}
  \affiliation{Jet Propulsion Laboratory, California Institute of Technology, Pasadena, CA 91109, USA}
  \affiliation{Division of Physics, Mathematics, and Astronomy, California Institute of Technology, Pasadena, CA 91125, USA}
\author[0000-0002-9225-9428]{A. Naidu}
  \affiliation{Department of Physics, McGill University, 3600 rue University, Montr\'eal, QC H3A 2T8, Canada}
  \affiliation{McGill Space Institute, McGill University, 3550 rue University, Montr\'eal, QC H3A 2A7, Canada}
\author[0000-0002-8912-0732]{A.~B. Pearlman}
  \altaffiliation{NDSEG Research Fellow.}
  \altaffiliation{NSF Graduate Research Fellow.}
  \affiliation{Division of Physics, Mathematics, and Astronomy, California Institute of Technology, Pasadena, CA 91125, USA}
\author{L. Spitler}
  \affiliation{Max Planck Institut f\"{u}r Radioastronomie, Auf dem H\"{u}gel 69, D-53121, Bonn, Germany}
\author[0000-0003-3772-2798]{K.~M. Bandura}
  \affiliation{CSEE, West Virginia University, Morgantown, WV 26505, USA}
  \affiliation{Center for Gravitational Waves and Cosmology, West Virginia University, Morgantown, WV 26505, USA}
\author[0000-0002-3615-3514]{M. Bhardwaj}
  \affiliation{Department of Physics, McGill University, 3600 rue University, Montr\'eal, QC H3A 2T8, Canada}
  \affiliation{McGill Space Institute, McGill University, 3550 rue University, Montr\'eal, QC H3A 2A7, Canada}
\author[0000-0003-2047-5276]{T. Cassanelli}
  \affiliation{Dunlap Institute for Astronomy \& Astrophysics, University of Toronto, 50 St.~George Street, Toronto, ON M5S 3H4, Canada}
  \affiliation{David A.~Dunlap Institute Department of Astronomy \& Astrophysics, University of Toronto, 50 St.~George Street, Toronto, ON M5S 3H4, Canada}
\author[0000-0002-3426-7606]{P. Chawla}
  \affiliation{Department of Physics, McGill University, 3600 rue University, Montr\'eal, QC H3A 2T8, Canada}
  \affiliation{McGill Space Institute, McGill University, 3550 rue University, Montr\'eal, QC H3A 2A7, Canada}
\author[0000-0002-3382-9558]{B.~M. Gaensler}
  \affiliation{Dunlap Institute for Astronomy \& Astrophysics, University of Toronto, 50 St.~George Street, Toronto, ON M5S 3H4, Canada}
  \affiliation{David A.~Dunlap Institute Department of Astronomy \& Astrophysics, University of Toronto, 50 St.~George Street, Toronto, ON M5S 3H4, Canada}
\author[0000-0003-1884-348X]{D.~C. Good}
  \affiliation{Department of Physics \& Astronomy, 6224 Agricultural Road, Vancouver, BC V6T 1Z1, Canada}
\author[0000-0003-3059-6223]{A. Josephy}
  \affiliation{Department of Physics, McGill University, 3600 rue University, Montr\'eal, QC H3A 2T8, Canada}
  \affiliation{McGill Space Institute, McGill University, 3550 rue University, Montr\'eal, QC H3A 2A7, Canada}
\author[0000-0002-5307-2919]{R. Karuppusamy}
  \affiliation{Max Planck Institut f\"{u}r Radioastronomie, Auf dem H\"{u}gel 69, D-53121, Bonn, Germany}
\author[0000-0002-5575-2774]{A. Keimpema}
  \affiliation{Joint Institute for VLBI ERIC, Oude Hoogeveensedijk 4, 7991PD Dwingeloo, The Netherlands}
\author[0000-0002-8139-8414]{A.~Yu. Kirichenko}
  \affiliation{Instituto de Astronomía, Universidad Nacional Autónoma de México, Apdo.~Postal 877, Ensenada, Baja California 22800, México}
  \affiliation{Ioffe Institute, 26 Politekhnicheskaya st., St.~Petersburg 194021, Russia}
\author[0000-0001-6664-8668]{F. Kirsten}
  \affiliation{Department of Space, Earth and Environment, Chalmers University of Technology, Onsala Space Observatory, 439 92, Onsala, Sweden}
\author[0000-0003-0249-7586]{J. Kocz}
  \affiliation{Division of Physics, Mathematics, and Astronomy, California Institute of Technology, Pasadena, CA 91125, USA}
\author[0000-0002-4209-7408]{C. Leung}
  \affiliation{MIT Kavli Institute for Astrophysics and Space Research, Massachusetts Institute of Technology, 77 Massachusetts Ave, Cambridge, MA 02139, USA}
  \affiliation{Department of Physics, Massachusetts Institute of Technology, 77 Massachusetts Ave, Cambridge, MA 02139, USA}
\author[0000-0001-9814-2354]{B. Marcote}
  \affiliation{Joint Institute for VLBI ERIC, Oude Hoogeveensedijk 4, 7991~PD Dwingeloo, The Netherlands}
\author[0000-0002-4279-6946]{K. Masui}
  \affiliation{MIT Kavli Institute for Astrophysics and Space Research, Massachusetts Institute of Technology, 77 Massachusetts Ave, Cambridge, MA 02139, USA}
  \affiliation{Department of Physics, Massachusetts Institute of Technology, 77 Massachusetts Ave, Cambridge, MA 02139, USA}
\author[0000-0002-0772-9326]{J. Mena-Parra}
  \affiliation{MIT Kavli Institute for Astrophysics and Space Research, Massachusetts Institute of Technology, 77 Massachusetts Ave, Cambridge, MA 02139, USA}
\author[0000-0003-2095-0380]{M. Merryfield}
  \affiliation{Department of Physics, McGill University, 3600 rue University, Montr\'eal, QC H3A 2T8, Canada}
  \affiliation{McGill Space Institute, McGill University, 3550 rue University, Montr\'eal, QC H3A 2A7, Canada}
\author[0000-0002-2551-7554]{D. Michilli}
  \affiliation{Department of Physics, McGill University, 3600 rue University, Montr\'eal, QC H3A 2T8, Canada}
  \affiliation{McGill Space Institute, McGill University, 3550 rue University, Montr\'eal, QC H3A 2A7, Canada}
\author[0000-0001-6898-0533]{C.~J. Naudet}
  \affiliation{Jet Propulsion Laboratory, California Institute of Technology, Pasadena, CA 91109, USA}
\author[0000-0003-0510-0740]{K. Nimmo}
  \affiliation{Anton Pannekoek Institute for Astronomy, University of Amsterdam, Science Park 904, 1098 XH, Amsterdam, The Netherlands}
  \affiliation{ASTRON, Netherlands Institute for Radio Astronomy, Oude Hoogeveensedijk 4, 7991 PD Dwingeloo, The Netherlands}
\author[0000-0002-4795-697X]{Z. Pleunis}
  \affiliation{Department of Physics, McGill University, 3600 rue University, Montr\'eal, QC H3A 2T8, Canada}
  \affiliation{McGill Space Institute, McGill University, 3550 rue University, Montr\'eal, QC H3A 2A7, Canada}
\author[0000-0002-8850-3627]{T.~A. Prince}
  \affiliation{Division of Physics, Mathematics, and Astronomy, California Institute of Technology, Pasadena, CA 91125, USA}
  \affiliation{Jet Propulsion Laboratory, California Institute of Technology, Pasadena, CA 91109, USA}
\author{M. Rafiei-Ravandi}
  \affiliation{Perimeter Institute for Theoretical Physics, 31 Caroline Street N, Waterloo ON N2L 2Y5, Canada}
\author[0000-0003-1842-6096]{M. Rahman}
  \affiliation{Dunlap Institute for Astronomy \& Astrophysics, University of Toronto, 50 St.~George Street, Toronto, ON M5S 3H4, Canada}
\author[0000-0002-6823-2073]{K. Shin}
  \affiliation{MIT Kavli Institute for Astrophysics and Space Research, Massachusetts Institute of Technology, 77 Massachusetts Ave, Cambridge, MA 02139, USA}
  \affiliation{Department of Physics, Massachusetts Institute of Technology, 77 Massachusetts Ave, Cambridge, MA 02139, USA}
\author{K.~M. Smith}
  \affiliation{Perimeter Institute for Theoretical Physics, 31 Caroline Street N, Waterloo ON N2L 2Y5, Canada}
\author[0000-0001-9784-8670]{I.~H. Stairs}
  \affiliation{Department of Physics \& Astronomy, 6224 Agricultural Road, Vancouver, BC V6T 1Z1, Canada}
\author[0000-0003-2548-2926]{S.~P. Tendulkar}
  \affiliation{Department of Physics, McGill University, 3600 rue University, Montr\'eal, QC H3A 2T8, Canada}
  \affiliation{McGill Space Institute, McGill University, 3550 rue University, Montr\'eal, QC H3A 2A7, Canada}
\author[0000-0003-4535-9378]{K. Vanderlinde}
  \affiliation{Dunlap Institute for Astronomy \& Astrophysics, University of Toronto, 50 St.~George Street, Toronto, ON M5S 3H4, Canada}
  \affiliation{David A.~Dunlap Institute Department of Astronomy \& Astrophysics, University of Toronto, 50 St.~George Street, Toronto, ON M5S 3H4, Canada}

\newcommand{\allacks}{

A.B.P. acknowledges support by the Department of Defense~(DoD) through the National Defense Science and Engineering Graduate~(NDSEG) Fellowship Program and by the National Science Foundation~(NSF) Graduate Research Fellowship under Grant~No.~\text{DGE-1144469}.
B.M. acknowledges support from the Spanish Ministerio de Econom\'ia y Competitividad (MINECO) under grant AYA2016-76012-C3-1-P.
D.M. is a Banting Fellow
F.K. is supported by the Swedish Research Council.
FRB research at UBC is supported by an NSERC Discovery Grant and by the Canadian Institute for Advanced Research.
J.W.T.H. acknowledges funding from an NWO Vici fellowship
L.G.S. is a Lise-Meitner independent research group leader and acknowledges support from the Max Planck Society. 
M.B. is supported by an FRQNT Doctoral Research Award.
P.C. is supported by an FRQNT Doctoral Research Award.
P.S. is a Dunlap Fellow and an NSERC Postdoctoral Fellow. 
B.M.G. acknowledges the support of the Natural Sciences and Engineering Research Council of Canada (NSERC) through grant RGPIN-2015-05948, and of the Canada Research Chairs program.
V.M.K. holds the Lorne Trottier Chair in Astrophysics \& Cosmology and a Canada Research Chair and receives support from an NSERC Discovery Grant and Herzberg Award, from an R. Howard Webster Foundation Fellowship from the Canadian Institute for Advanced Research (CIFAR), and from the FRQNT Centre de Recherche en Astrophysique du Quebec.
W.A.M, T.A.P, and C.J.N acknowledge support by the Jet Propulsion Laboratory's Spontaneous Concept Research and Technology Development program.
Z.P. is supported by a Schulich Graduate Fellowship.

}

\begin{abstract}
We report on simultaneous radio and X-ray observations of the repeating fast radio burst source \src\ using the 
Canadian Hydrogen Intensity Mapping Experiment (CHIME), Effelsberg, and Deep Space Network (DSS-14 and DSS-63) radio telescopes and the \cxofull. During 33\,ks of \chandra\ observations, we detect no radio bursts in overlapping Effelsberg or Deep Space Network observations and a single radio burst during CHIME/FRB source transits. We detect no X-ray events in excess of the background during the \cxo\ observations. These non-detections imply a 5-$\sigma$ limit of $<5\times10^{-10}$\,\flucgs\ for the 0.5--10\,keV fluence of prompt emission at the time of the radio burst and $1.3\times10^{-9}$\,\flucgs\ at any time during the \chandra\ observations at the position of \src. Given the host-galaxy redshift of \src\ ($z\sim0.034$), these correspond to energy limits of $<1.6\times10^{45}$\,erg and $<4\times10^{45}$\,erg, respectively. We also place a 5-$\sigma$ limit of $<8\times10^{-15}$\,\fluxcgs\ on the 0.5--10\,keV absorbed flux of a persistent source at the location of \src. This corresponds to a luminosity limit of $<2\times10^{40}$\,\lumcgs. Using \fermi/GBM data we search for prompt gamma-ray emission at the time of radio bursts from \src\ and find no 
significant bursts, placing a limit of $4\times10^{-9}$\,\flucgs\ on the 10--100\,keV fluence. We also search \fermi/LAT data for periodic modulation of the gamma-ray brightness at the 16.35-day period of radio-burst activity and detect no significant modulation. We compare these deep limits to the predictions of various fast radio burst models, but conclude that similar X-ray constraints on a closer fast radio burst source would be needed to strongly constrain theory.
\end{abstract}

\section{Introduction}

Fast radio bursts (FRBs) are a new class of radio
transient with unknown origins \citep[see][for reviews]{cc19,phl19}.  
They are millisecond-long, bright (peak flux densities $\sim0.1$--10\,Jy at $\sim1$\,GHz) bursts {\ps and have been observed at frequencies from 300\,MHz \citep{cab+20} to 8\,GHz \citep{gsp+18}}.
Their distances, both based on their dispersion measure (DM) excesses 
\citep[in comparison to the expected Milky Way contributions;][]{ne2001,ymw17} 
and measured host-galaxy redshifts for a few sources \citep{clw+17,bdp+19,rcd+19,pmm+19,mnh+20}, 
are extragalactic, and the most distant
sources appear to come from cosmological distances \citep[i.e., $z\gapp 0.5$;][]{tsb+13}.
The extreme luminosities and short duration of FRBs point to coherent emission
originating from a compact object. Prior to the discovery of repeat bursts from 
some FRB sources, most models invoked cataclysmic phenomena to explain the extreme
energetics of FRBs \citep[for a catalog of models, see][]{pww+18}.
However, since the discovery of repeat bursts from \repeater\ \citep{ssh+16a},
models that can account for repetition have become increasingly the focus of theoretical work.

One central engine in particular has garnered a lot of attention: the millisecond magnetar.
In this model, an FRB is powered by a young, recently formed millisecond magnetar \citep[e.g.,][]{lyu14,bel17,mbm17} and may have a high-energy counterpart.
The older, much less energetic, magnetars in our Galaxy are known to power X-ray and gamma-ray bursts and
flares on timescales of milliseconds to seconds \citep[see][for a review]{kb17}, which are similar to the duration 
of FRBs. The high-energy burst emission of magnetars comes in at least two classes: 
giant flares and short X-ray bursts. To date, only three magnetar giant flares have been
detected in our Galaxy \citep{ekl+80,hkm+99,hbs+05} with X-ray peak luminosities in the
range $\sim10^{44}-10^{47}$\,\lumcgs. Short X-ray bursts from magnetars are emitted much
more frequently but are much fainter than giant flares 
\citep[peak X-ray luminosities of $\sim10^{36}-10^{43}$\,\lumcgs; e.g.,][]{gwk+99,gwk+00,sk11}.

\citet{sbh+17} undertook several campaigns of coordinated X-ray and radio observations of \repeater, to probe for coincident high-energy emission during the radio bursts. With these observations, upper limits were placed on X-ray (0.5--10\,keV) and gamma-ray (10--100\,keV) emission at the time of radio bursts. Owing to the relatively large distance to \repeater\ ($z\sim0.193$; luminosity distance of 972\,Mpc), these limits were found to be $\sim10\times$ above what is expected for a magnetar giant flare \citep{sbh+17}.

The recent success of the CHIME/FRB Collaboration in discovering repeating FRBs has led
to several sources that could be much closer than \repeater, based on their low DM excesses \citep{abb+19c}.
One of these sources, \src, was subsequently localized with milliarcsecond precision 
to a spiral galaxy at $z=0.0337\pm0.0002$ (luminosity distance of 149\,Mpc)
using observations from the European VLBI Network \citep{mnh+20}.
Recently, a 16.35-day periodicity in the burst activity of \src\ was found, where
the source seems to be active in a $\sim5$ day window \citep{aab+20}, although an aliased, shorter period cannot presently be excluded.
Armed with this localization, and knowledge of the periodic activity level, we were able to perform a 
deep, targeted, search for  X-ray emission using the {\em Chandra X-ray Observatory} 
coordinated with radio observations at times when the detection of radio bursts from 
the source were highly probable.
The greater proximity
of \src\ compared to \repeater\ allows us to probe 
$\sim40\times$ deeper in energy for such emission.
Previously, limits have been placed on the high-energy emission of \src\ during its active
phases using {\ps {\em INTEGRAL} \citep{psf+20}, \swift/XRT \citep{tvc+20}, and \cxo\ \citep{klhy20}}\footnote{based on the same \cxo\ observations presented here.} {\ps Other
studies have also placed limits on the gamma-ray emission of a large sample of FRB sources \citep[e.g.,][]{tkp16,ccb+19}.}

Here we present simultaneous deep X-ray and radio observations on
2019 December 3 and 18 performed with the goal of detecting or constraining any
X-ray counterparts to the radio bursts from \src. We also present a search for
gamma-ray emission at the times of radio bursts from \src.
We describe the \chandra\ (X-ray), \fermi\ (gamma-ray), Effelsberg, Deep Space Network, and CHIME (radio) observations in Section \ref{sec:obs}.
In Section \ref{sec:results} we
present the results of our search for bursts in the radio observations
as well as X-ray (\cxo) and gamma-ray (\fermi) emission 
both at the time of radio bursts and at anytime during the high-energy
observations. We discuss the 
significance of these results in Section \ref{sec:discussion}.

\section{Observations}
\label{sec:obs}

\begin{figure}
\includegraphics[width=\columnwidth]{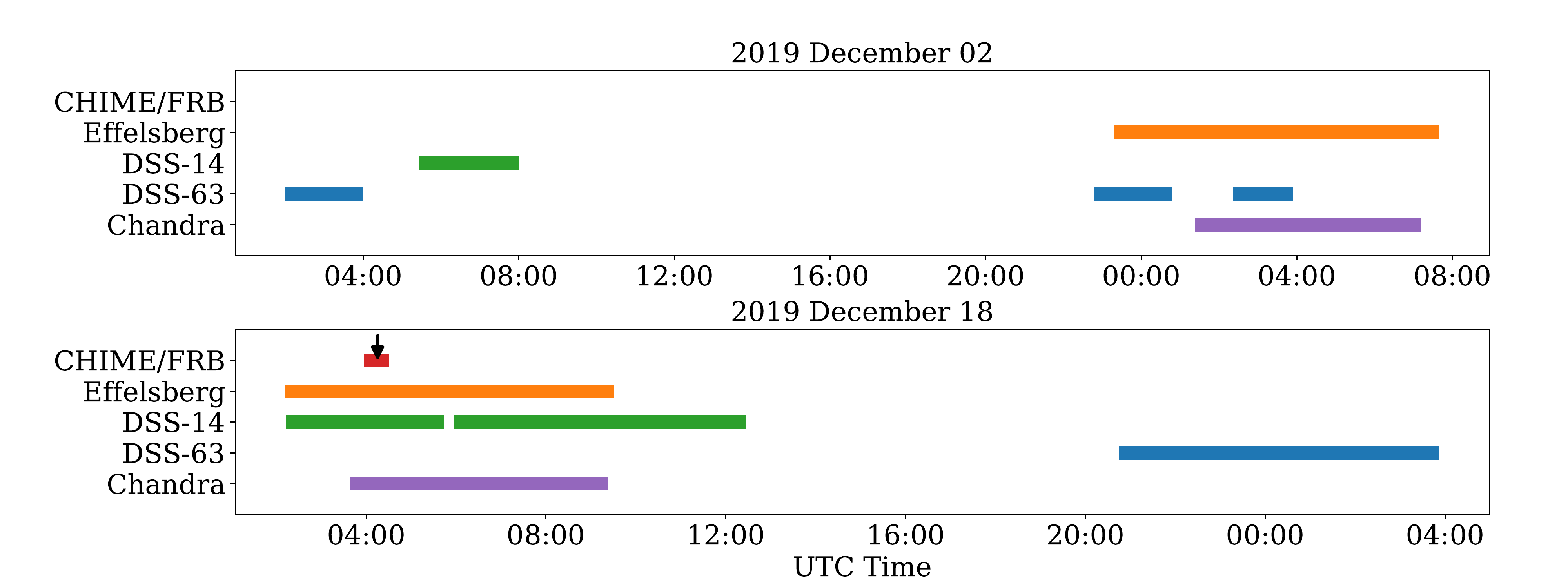}
\figcaption{\ps
Timeline of \cxo\ observations (purple) and the coordinated radio observations from CHIME/FRB (red), Effelsberg (orange), and the Deep Space Network (DSS-14, in green, and DSS-63, in blue, telescopes). The bars show the times when each telescope was observing \src. The arrow on 2019 December 18 marks the time of the CHIME/FRB-detected burst. \label{fig:obs}
}
\end{figure}

\floattable
\begin{deluxetable*}{ccccc} 
\tabletypesize{\normalsize} 
\tablecolumns{5} 
\tablewidth{0pt} 
\tablecaption{ Summary of Joint X-ray/Radio Observations \label{tab:obs}} 
\tablehead{ 
\colhead{Telescope} & \colhead{Obs ID/}    & \colhead{Start time}    & \colhead{End time} &  \colhead{Exposure time}\\ 
\colhead{}          & \colhead{Frequency (MHz)} & \colhead{(UTC)}  & \colhead{(UTC)}  & \colhead{(s)}}
\startdata 
\multirow{2}{*}{\cxo} & 23081      & 2019-12-03 01:33:03 & 2019-12-03 07:01:53 & 16390 \\
                      & 23082      & 2019-12-18 03:47:29 & 2019-12-18 09:13:36 & 16300 \\
\hline
\multirow{1}{*}{CHIME/FRB\tablenotemark{a}}& 400--800    & 2019-12-18 04:06:36 & 2019-12-18 04:21:22 & 886 \\ 
\hline
\multirow{2}{*}{Effelsberg} & \multirow{2}{*}{1210--1510}  &   2019-12-02 23:29:27 & 2019-12-03 07:29:27 & 28800 \\
                            &   &   2019-12-18 02:21:17 & 2019-12-18 09:21:17 & 25200 \\
\hline
\multirow{8}{*}{DSS-14} & \multirow{8}{*}{1360--1720} & 2019-12-02 05:37:28 & 2019-12-02 07:01:28 & 5040  \\
                        &  & 2019-12-02 07:09:02 & 2019-12-02 07:50:16 & 2474  \\
                        &  & 2019-12-18 02:22:04 & 2019-12-18 03:51:04 & 5340  \\
                        &  & 2019-12-18 04:01:12 & 2019-12-18 05:35:12 & 5640  \\
                        &  & 2019-12-18 06:05:42 & 2019-12-18 07:22:42 & 4620  \\
                        &  & 2019-12-18 07:30:16 & 2019-12-18 09:04:16 & 5640  \\
                        &  & 2019-12-18 09:11:46 & 2019-12-18 10:45:46 & 5640  \\
                        &  & 2019-12-18 10:53:32 & 2019-12-18 12:18:32 & 5100  \\
\hline
\multirow{7}{*}{DSS-63} & & 2019-12-02 02:10:34 & 2019-12-02 03:49:34 & 5940  \\
                        & & 2019-12-02 22:58:40 & 2019-12-03 00:37:40 & 5940  \\
                        &\multirow{2}{*}{2205--2310 \&} & 2019-12-03 02:32:46 & 2019-12-03 03:43:29 & 4243  \\
                        &\multirow{2}{*}{8180--8575} & 2019-12-18 20:53:48 & 2019-12-18 22:27:48 & 5640  \\
                        & & 2019-12-18 22:36:30 & 2019-12-19 00:10:30 & 5640  \\
                        & & 2019-12-19 00:18:26 & 2019-12-19 01:52:26 & 5640  \\
                        & & 2019-12-19 02:00:34 & 2019-12-19 03:43:14 & 6120 \\ 
\enddata 
\tablenotetext{a}{Start, end, and exposure times based on time spent by source within the 600\,MHz FWHM of the CHIME/FRB formed beams.}
\end{deluxetable*}

\subsection{Chandra X-ray Observatory}

\src\ was observed by \chandra\ on 2019 December 3 (ObsID 23081) and 
2019 December 18 (ObsID 23082) at epochs consistent with the ``on-phase'' of
the periodic activity of \src\ identified by \citet{aab+20}.
The ACIS-S3 detector was operated in VFAINT mode with a 1/8 sub-array read out 
providing a $8'\times1'$ field of view and a 0.4-s frame time. {\ps The exposure times were both $\sim16$\,ks, as listed in 
Table \ref{tab:obs}}. {\ps Figure \ref{fig:obs} shows a timeline of the \cxo\ observations and how they overlap with radio observations.}

The resulting data were analyzed using 
CIAO\footnote{Chandra Interactive Analysis of Observations. 
\url{http://cxc.harvard.edu/ciao/}} version 
4.12 \citep{fma+06} following standard procedures recommended by 
the \cxo\ X-ray Center.
Source events were extracted from a 1\arcsec-radius region (95\% encircled energy)
centered on the position of \src\ and arrival times were corrected to 
the Solar-System Barycenter using the source position measured by \citet{mnh+20} to a precision of $\sim2$\,milliarcseconds with the European VLBI Network (EVN).

\subsection{CHIME/FRB}

 The CHIME/FRB backend continuously searches total-intensity, polarization-summed time series from each of the 1,024 beams formed across CHIME's $2\degrees\times120\degrees$ field-of-view. 
 The time series have a 0.98304-ms time resolution and 16,384 frequency channels across the 400--800 MHz band. 
 The backend uses real-time radio-frequency interference (RFI) mitigation and a tree dedispersion algorithm to search over a wide range of trial DMs. Dispersed signals with integrated S/N values greater than the system's configurable threshold are forwarded to a post-detection pipeline to classify sources as RFI, known Galactic sources, or unknown Galactic or extragalactic signals (by comparing to predicted Galactic contributions to DM). Signals are classified as FRBs (i.e., unknown extragalactic) if they are not associated with any known Galactic sources, and their observed DMs exceed the maximum values predicted by Galactic DM models \citep{ne2001,ymw17}.
 See \citet{abb+18} for a detailed description of the CHIME/FRB system.
 
 On 2019 December 3, CHIME was offline for upgrades and so was unable to search for bursts at that time. 
 On 2019 December 18, \src\ was within the FWHM (at 600\,MHz) of the CHIME/FRB beams for 14.7 min (see Table \ref{tab:obs}).
 During that period, the source moved through the four columns of synthesized beams.  As such, the sensitivity to
 \src\ varied significantly over the course of the transit. 

\subsection{Effelsberg Radio Telescope}
The Effelsberg 100-m radio telescope observed \rthree\ with the 7-beam receiver (P217mm) at a center frequency of 1.36 GHz. The central beam was pointed at the precise position measured by the EVN localization \citep{mnh+20}. {\ps These observations spanned the full extent of the \cxo\ observations (see Figure \ref{fig:obs}) and their start times, end times, and total on-source time are given in Table~\ref{tab:obs}}. The PFFTS digital backend recorded total intensity spectral data with a time resolution of 54.6 $\upmu$s, 512 frequency channels, and a bandwidth of 300 MHz ($\Delta \nu$ = 0.586 MHz). Before processing, the PFFTS data  were converted from 32-bit floats to 8-bit unsigned integers in {\tt sigproc} filterbank format. 

The data were searched using the {\tt PRESTO} search software \citep{presto}\footnote{\url{https://github.com/scottransom/presto}}. Broadband, impulsive RFI was removed using an algorithm that first re-scales each frequency channel according to the standard deviation and median of that channel and then calculates a zero-DM timeseries. Statistically anomalous time samples were identified by applying an S/N threshold, and values for each frequency channel in that time sample were replaced with Gaussian noise with the statistics of that channel. The cleaned filterbank was then passed to {\tt rfifind} for further RFI excision. The data were then downsampled by a factor of eight in time and dedispersed with 100 trial DMs ranging from 300\dmunits\ to 400\dmunits\ \citep[\src\ has a DM of 349\,\dmunits;][]{abb+19c} in steps of 1\dmunits\ with {\tt prepsubband}. Each time series was convolved with a template bank of boxcar matched filters yielding effective time resolutions of 0.44~ms to 17.5 ms, and candidate bursts were identified in each timeseries by applying a detection threshold of S/N~$> 6$ ({\tt single\_pulse\_search.py}). The results were inspected by eye, and promising candidates were further investigated by looking at a time-frequency snapshot around each candidate. 

\subsection{Deep Space Network}

The Deep Space Network (DSN) observed \src\ for a total of $\sim 22$\,hr, {\ps partially overlapping with the \cxo\ observations (see Figure \ref{fig:obs}),} using \mbox{DSS-14} and DSS-63, two 70-m diameter radio antennas located in Goldstone, California and Robledo, Spain. \src\ was observed at $L$-band (center frequency of $1.5$\,GHz; data recorded in left circular polarization) using DSS-14 for a total of 11\,hr over eight separate scans. DSS-63 observed \src\ simultaneously at $S$-band (center frequency of $2.3$\,GHz) and $X$-band (center frequency of $8.4$\,GHz) with data recorded in both left and right circular polarization for a total of 11\,hr in seven separate scans (see Table \ref{tab:obs}). The $L$-band system on DSS-14 spans roughly 500\,MHz of bandwidth, but only $250$\,MHz of the total bandwidth was usable during our observations after RFI mitigation. The data at $S$-band and $X$-band were recorded with bandwidths of $105$\,MHz and $395$\,MHz, respectively.

Data were recorded using pulsar backends that record channelized power spectral density measurements in filterbank format. The $L$-band data were recorded with a time and frequency resolution of 102.4\,$\upmu$s and 0.625\,MHz, respectively. The $S$-band and $X$-band data were recorded with a time and frequency resolution of $2.2$\,ms and $0.464$\,MHz, respectively. We performed short observations of a bright pulsar (PSR~B0329+54) at various times throughout the observing campaign to validate the quality of the data. The data were flux calibrated by measuring the $T_{\text{sys}}$ at each frequency band while the antenna was in the stow position. We then corrected the $T_{\text{sys}}$ values for elevation effects, which were minimal since all of our observations occurred when the source elevation was above 20$^{\circ}$.

The data processing procedures followed those described in previous DSN studies of pulsars (e.g.,~\citealt{mpd+17, pmp+18, pmp19}). In each data set, we corrected for the bandpass slope across the frequency band. Bad frequency channels corrupted by RFI were identified using the PSRCHIVE software~\citep{hvm+04} and masked. We also subtracted the moving average from each data point using 0.5\,s around each time sample in order to remove any long timescale temporal variability. The cleaned data from each epoch were then dedispersed with trial DMs between 300 and 400\,\dmunits. We searched for FRBs using a matched filtering algorithm, where each dedispersed time-series was convolved with logarithmically spaced boxcar functions with widths ranging between 1--300 times the native time resolution. FRB candidates with detection S/N$>$6 were saved and classified using a GPU-accelerated machine learning pipeline based on the \texttt{FETCH} (Fast Extragalactic Transient Candidate Hunter) package~\citep{aab+19}.

\subsection{Fermi Gamma Ray Space Telescope}

The \fermi\ telescope has two sets of detectors on board, the Gamma-ray Burst Monitor \citep[GBM;][]{mlb+09} and the Large Area Telescope \citep[LAT;][]{aaa+09}.
The GBM consists of 12 sodium iodide (NaI; 8 keV -- 40 MeV) and 2 bismuth germanate (BGO; 300 keV -- 40 MeV) scintillators pointed in various directions to provide all-sky coverage to gamma-rays. In this work we use only the NaI detectors.
The GBM instrument records data in several different data products, but here we use only the time-tagged events (TTE) data which provides event data with 2-$\upmu$s time resolution and 128 energy channels. 
The LAT is a pair-conversion telescope providing sensitivity to gamma-ray photons in the range 20 MeV--300\,GeV in a 2.4 sr (20\% of sky) field of view. The LAT images the sky with a time resolution of 10 $\upmu$s or better.
The LAT collaboration periodically releases improved reprocessing of their gamma-ray events. Here we use the most recent release, Pass 8.

\section{Analysis and Results}
\label{sec:results}
\subsection{Radio Bursts}
\label{sec:radiobursts}

During the 2019 December 18 transit of \src\ over CHIME, which was simultaneous with a
\cxo\ observation, 
a single radio burst was detected by CHIME/FRB. The burst was detected at MJD 58835.17721035 (barycentric {\ps after correcting for dispersive delay}), {\ps 446\,s after the start of the \cxo\ observation,} with a band-averaged S/N of 12.8 which corresponds to a peak flux density 
of $0.4\pm0.2$\,Jy and fluence of $2.9\pm0.7$\,Jy\,ms \citep[see][for additional details on this burst]{aab+20}.

In the simultaneous Effelsberg observations, no bursts with S/N $>$ 6 were identified by the {\tt PRESTO} search.  Assuming a system equivalent flux density of 20~Jy for the P217mm receiver and S/N $>$ 6, the fluence threshold is 0.15~Jy~ms $\sqrt{(w/\mbox{1 ms})}$, where $w$ is the burst duration in ms.
The Effelsberg time series were also manually inspected around the time of detected CHIME/FRB bursts and no excess was found. 
In the DSN observations listed in Table \ref{tab:obs}, no radio bursts were detected. For a pulse width of $w$, the fluence thresholds (for S/N $>$ 6) on the peak flux densities during these epochs are:
0.25~Jy~ms $\sqrt{(w/\mbox{1 ms})}$ at $L$-band,
0.29~Jy~ms $\sqrt{(w/\mbox{1 ms})}$ at $S$-band,
0.14~Jy~ms $\sqrt{(w/\mbox{1 ms})}$ at $X$-band.

\subsection{Limits on Prompt X-ray Emission}
\label{sec:cxobursts}

We searched the \cxo\ observations both for X-ray photons arriving nearby in time to the CHIME/FRB-detected radio burst and at anytime during the observations.
In the 2019 December 3 \cxo\ observation, a single photon was detected at the source position, but there were no detected radio bursts in overlapping radio observations.
In the 2019 December 18 \cxo\ observation, a single photon was detected at the source position, {\ps 4.7\,hr after the CHIME/FRB-detected radio burst and 500\,s before the end of the \cxo\ observation}. We take into account the dispersion delay of the radio bursts (9\,s at 400 MHz) when comparing to the times of high-energy photons. 
The background count rate {\ps in the source extraction region }during the observation was $6\times10^{-5}$ counts\,s$^{-1}$.
{\ps This leads to a probability of $64\%$ of detecting one or more photons within
4.7\,hr of the radio burst.}
Given this high false alarm probability, we have no
reason to associate the detection with \src.
For both observations, the detection of a single X-ray count within the source
extraction region of \src\ is consistent with the background count rate.

{\ps Following \citet{sbh+17}, we place upper limits using Poisson statistics and the Bayesian method of \citet{kbn91}. For all limits in this work, we use a stringent confidence level of 0.9999994, the equivalent of the 5-$\sigma$ width of a Gaussian distribution. For brevity, we refer to this confidence level as ``5-$\sigma$'' below.
We first derive a ``model-independent'' limit, that is, assuming an equal probability
of a source photon occurring across the 0.5--10\,keV band (note that this is effectively
assuming a flat spectral model with zero X-ray absorption; see below for exploration of more reasonable models).
This 5-$\sigma$ confidence upper 
limit on the 0.5--10\,keV fluence for a single X-ray burst at the time
of the detected radio bursts is $5\times10^{-10}$\,erg\,cm$^{-2}$
corresponding to $1.6\times10^{45}$\,erg at the luminosity distance of \src.}
These fluence and energy limits are valid for any burst duration contained within 446\,s before the radio bursts (i.e., from the beginning of the observation) and 4.7\,hr after (i.e., up to the time of the \chandra\ background photon).
The fluence limit for an X-ray burst arriving at any other time during the \cxo\ observations is $1.3\times10^{-9}$\,erg\,cm$^{-2}$ for an assumed duration of 5\,ms,
corresponding to an energy limit of $4\times10^{45}$\,erg.

As discussed in \citet{sbh+17}, the implied limit on the emitted energy
of a putative X-ray burst depends strongly on the underlying spectral model of the
burst. By assuming a spectral model and taking into account the spectral response of 
\cxo, a fluence limit for that underlying spectral model  can be calculated. 
To generate the assumed source spectra we
used {\tt XSPEC} v12.10.1f with abundances from \citet{wam00} and photoelectric 
cross-sections from \citet{vfky96}. In order to enable direct comparison, we assume
the same fiducial models used by \citet{sbh+17} for \repeater:
a blackbody spectrum with $kT=10$\,keV as observed in magnetar hard X-ray bursts \citep[e.g.,][]{lgb+12,akb+14},
a cutoff power-law with index $\Gamma=0.5$ and cutoff energy of 500\,keV, similar to a SGR~1806$-$20-like giant flare 
spectrum \citep{mca+05,pbg+05} and a power-law model with index $\Gamma=2$ as an example soft
spectrum, a contrast to the hard magnetar burst models. In Table \ref{tab:lims} 
we show the resulting fluence and energy limits assuming these source models.
For X-ray absorption, we assume two values, $10^{22}$\nhunits\ and $10^{24}$\nhunits. 
The first is a typical value for a sightline passing through the Milky Way and the disk 
of a Milky-Way-like host galaxy and the second is an extreme value to show the effects of a high degree of absorption
from material close to the source {\ps \citep[such as the surrounding supernova ejecta in the magnetar model;][]{mbm17}.}

\floattable
\begin{deluxetable}{ccccccccc} 
\tabletypesize{\normalsize} 
\tablecolumns{5} 
\tablewidth{0pt} 
\tablecaption{ Burst limits from \cxo\ for different X-ray spectral models \label{tab:lims}}
\tablehead{ 
\colhead{Model} & \colhead{\nh}  & \colhead{kT/$\Gamma$}   & \colhead{Absorbed 0.5--10 keV}  & \colhead{Unabsorbed 0.5--10 keV} & Extrapolated 10 keV--1 MeV \\ 
\colhead{}      & \colhead{(\nhunits)} & \colhead{(keV/-)} & \colhead{Fluence Limit}         & \colhead{Energy Limit\tablenotemark{a} }           & Energy Limit\tablenotemark{a} \\    
\colhead{}      & \colhead{}     & \colhead{}              & \colhead{($10^{-11}$\,\flucgs)} & \colhead{($10^{45}$\,erg)}  & \colhead{($10^{47}$\,erg)} } 
\startdata 
Blackbody & $10^{22}$ & 10  & $90$  & 3   & $0.7$ \\ 
Blackbody & $10^{24}$ & 10  & $200$ & 20 & $7$\tablenotemark{b} \\ 
Cutoff PL & $10^{22}$ & 0.5 & $50$  & 1.4   & $5$ \\ 
Cutoff PL & $10^{24}$ & 0.5 & $180$ & 20 & $90$\tablenotemark{b} \\ 
Soft PL   & $10^{22}$ & 2   & 20  & 0.9   & 0.014 \\ 
Soft PL   & $10^{24}$ & 2   & 120    & 50 & 0.8 \\ 
\enddata 
\tablenotetext{a}{Assuming the measured luminosity distance to \src, 149\,Mpc \citep{mnh+20}.}
\tablenotetext{b}{More stringent limits on these models are available from \fermi/GBM. See Section \ref{sec:gammalim}.}
\tablecomments{5-$\sigma$ confidence upper limits. See Section \ref{sec:cxobursts} for details.}
\end{deluxetable}

\subsection{Limits on Persistent X-ray Emission}
\label{sec:cxopointsource}

To place the best-possible limits on a persistent source we combined the two 
\cxo\ observations for a total of 33\,ks of exposure time. 
In these two observations, only two events were detected in a 
$1\arcsec$-radius region centered on the position of \src. 
We measure a 0.5--10\,keV background count rate in a $25\arcsec$-radius region chosen
to be away from the source of 0.7 counts s$^{-1}$ sq.\,arcsec$^{-2}$.
Given this background rate, the two detected counts are consistent with the 
background in the combined observations.
Using these detected and measured background rates, we measure a 5-$\sigma$
count rate limit of $5.5\times10^{-4}$\,counts\,s$^{-1}$, using the Bayesian method 
of \citet{kbn91}. Assuming a photoelectrically absorbed power-law 
source spectrum with $\Gamma=2$ and $\nh\sim1\times10^{22}$\,cm$^{-2}$, 
the 5-$\sigma$ upper limit on the persistent 0.5--10\,keV X-ray 
absorbed flux from \src\ or its host galaxy is $8\times10^{-15}$ erg cm$^{-2}$ s$^{-1}$. 
At the luminosity distance of \src\ this corresponds to an isotropic luminosity limit of
$2\times10^{40}$ erg s$^{-1}$.

\subsection{Limits on Prompt Gamma-ray Emission}
\label{sec:gammalim}

We searched data from the \fermi/GBM
for gamma-ray counterparts at the time of radio bursts from 
\src\ using a similar analysis to that in \citet{sbh+17}.
We searched the TTE GBM data in the energy range 10--100\,keV for
NaI detectors that were pointed $< 60\degrees$ from the source position.
The 2018 December 18 bursts in this work were not visible to GBM as the source was occulted by the Earth at the time.
However, of the 28 bursts in \citet{aab+20}, 12 bursts occurred at a time when TTE data were 
available and the source was $< 60\degrees$ from at least one NaI detector
and not occulted by the Earth.
For these bursts, we searched each TTE timeseries for excess counts in 1- and 5-ms bins 
in a 20-s window centered on the arrival time of the CHIME/FRB detected radio burst {\ps (after correcting for the dispersive delay)}. 
We find no signals that are not attributable to Poisson 
fluctuations from the background count rate at a 5-$\sigma$ confidence level.
Taking into account the effective area of the NaI detectors\footnote{Generated using the GBM Response Generator \url{https://fermi.gsfc.nasa.gov/ssc/data/analysis/rmfit/DOCUMENTATION.html}}
towards the source position at 
the time of each event, the background count rate, and assuming a burst timescale of 0.1\,s, 
we place an upper limit of $2\times10^{-8}$\,erg\,cm$^{-2}$ on the 10--100\,keV fluence. 
This corresponds to a 10--100\,keV burst isotropic energy limit of $6\times10^{46}$\,erg at the measured 
luminosity distance of \src. If we assume a burst of gamma-rays is emitted
at the time of each radio burst, the limit becomes $4\times10^{-9}$\,erg\,cm$^{-2}$. At the measured luminosity distance of \src, 
this corresponds to a 10--100\,keV burst energy limit of $1\times10^{46}$\,erg.
These limits are more constraining than the extrapolated limits for prompt emission from the \cxo\ observations presented
in Table \ref{tab:lims} for the highly-absorbed hard (10\,keV blackbody and cut-off power-law) models.
For those fiducial models the 10\,keV to 1\,MeV energy limits are $7\times10^{46}$\,erg and $1\times10^{48}$\,erg, respectively.
No bursts from this work or \citet{aab+20} occurred in the \fermi/LAT field-of-view.

\subsection{Search for Periodic Gamma-ray Emission}

All \fermi/LAT photons with energies above 1 GeV and within a $5^\circ$ radius region around the coordinates of the source were selected, conservatively reflecting the $\sim 3^{\circ}$ 95\% containment radius for the point spread function at 1 GeV. We then filtered the data based on event class and zenith angle to ensure data quality and exclude Earth-limb photons. This data spans all 11 years from {\ps MJD~54683 to MJD~58907}. We removed data outside of the Good Time Intervals and corrected for exposure in each phase bin, before folding the data at the measured 16.35-day period \citep{aab+20}. 
{\ps We performed an H-test \citep{drs89} on the resultant pulse profile and find no significant signal, with a false-alarm probability of 31.3\%.}

\section{Discussion}
\label{sec:discussion}

\subsection{Comparison to Previous Limits}
\label{sec:compprev}

\begin{figure}
\includegraphics[width=\columnwidth]{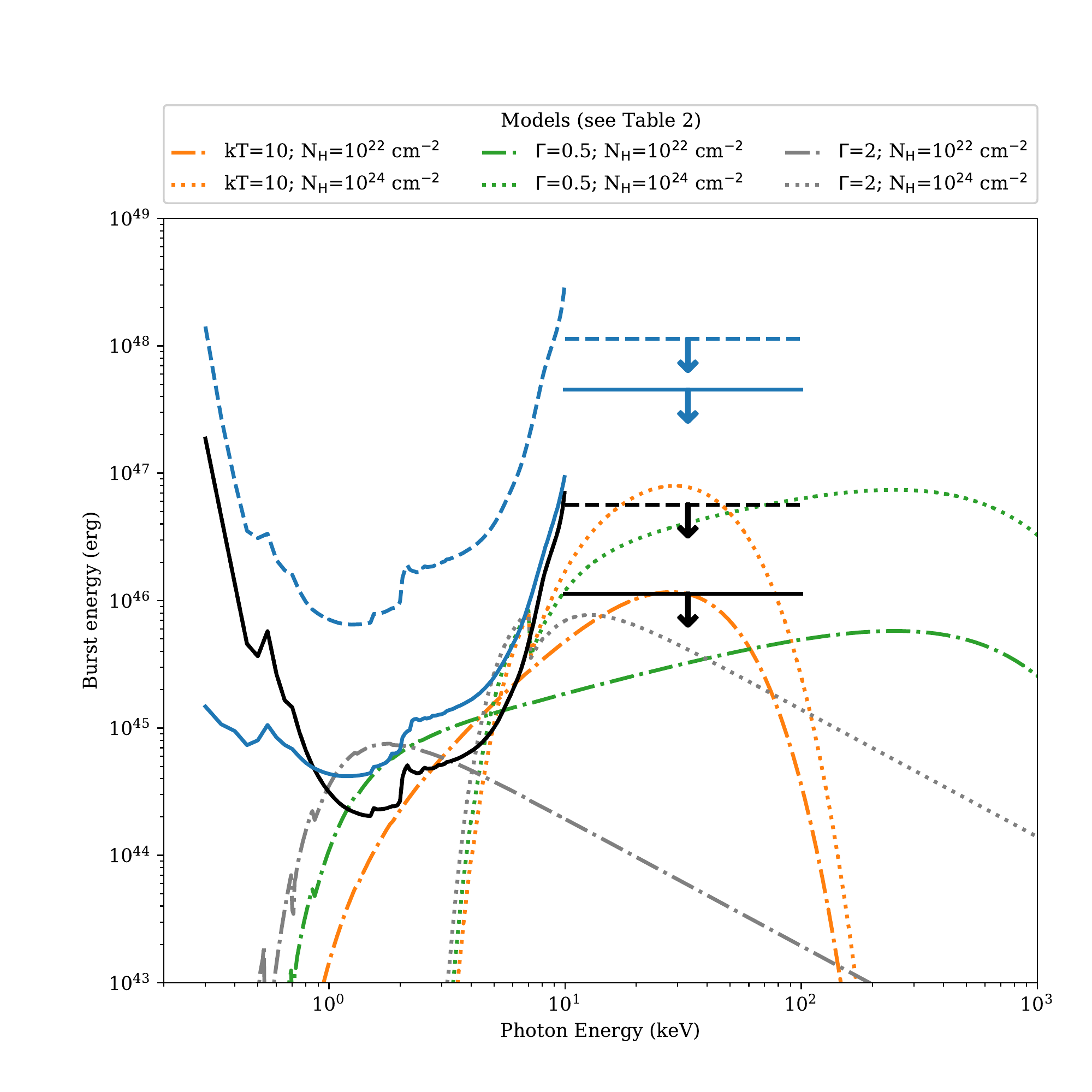}
\figcaption{
Limits on the energy of X-ray and gamma-ray bursts at the time of radio bursts from \src\ (in black; this work) and \repeater\ \citep[in blue; from][]{sbh+17}. The limits in the 0.5--10\,keV range are from \cxo, and in the 10--100\,keV range are from \fermi/GBM.
Dashed and solid lines show the 5-$\sigma$ upper limits as a function of X-ray photon energy, at the time of a single radio
burst and stacking those limits (see Section \ref{sec:compprev}), respectively.
The dot-dashed lines show different burst spectra that are photoelectrically
absorbed, assuming $\nh=10^{22}$\,cm$^{-2}$, plotted at their 0.5--10\,keV fluence 
limits that result from a stacked search of the times of the radio bursts.
The dotted lines show the same spectral models but with \nh=$10^{24}$\,cm$^{-2}$ 
to show the effects of possible heavy absorption local to the source. Orange lines represent a blackbody model 
with $kT=10$\,keV, green curves shows a cutoff power-law model with $\Gamma=0.5$ 
and $E_\mathrm{cut}=500$\,keV, and the grey curves show a soft power-law with
$\Gamma=2$ in order to illustrate how different underlying 
spectra affect the interpretation of the X-ray observations.
\label{fig:limits}
}
\end{figure}

The limits determined here can be compared to the similar 
campaign performed for \repeater\ using \xmm\ and \cxo\ observations that were simultaneous with radio observations \citep{sbh+17}. Figure \ref{fig:limits} shows the limits, in burst energy, as a function of photon energy for
both \repeater, from \citet{sbh+17}, and \src, from this work. As \src\ is 
$6.5$ times closer than 
\repeater, the single-burst energy limits from \cxo\ ACIS and \fermi\ GBM observations are $\sim40\times$
more constraining. However, the campaign on \repeater\ included several \cxo\ and \xmm\ observations during which 11 radio bursts
were detected, compared to the single burst detected for \src\ in this work.
This means that the {\ps (flat-model)} single-burst 0.5--10 keV energy limit for prompt emission from \src, $1.6\times10^{45}$\,erg, is only $\sim3\times$ more constraining than the combined limit for \repeater, $4\times10^{45}$\,erg, which was derived under the assumption that an X-ray burst of similar fluence was emitted near the time of each radio burst.

The \nh\ values assumed in the above calculations are the same as those taken
for \repeater\ \citep{sbh+17}, but
may not be applicable for \src.
From the DM budget presented by \citet{mnh+20} and the DM--\nh\ relation from \citet{hnk+13},
we can estimate what the \nh\ towards \src\ could be. The total DM measured for \src\ is 
349\dmunits.  Assuming the intergalactic medium (IGM) does not contribute significantly to \nh, we subtract the 
IGM contribution to the DM, determined from the DM--$z$ relation \citep{ino04}, 34\dmunits.
This leaves a Milky Way plus host DM of 291\dmunits, which from the DM--\nh\ relation roughly 
corresponds to $\nh=10^{22}$\,cm$^{-2}$, as used above. The high \nh\ value, $10^{24}$\,cm$^{-2}$, was used in \citet{sbh+17} to simulate extreme X-ray absorption local to the 
source due to a high ratio of atomic metals to free electrons, which could occur in a 
decades-old supernova remnant \citep{mbm17}. However, \citet{cab+20} argues against such a young
remnant for \src\ because of their recent detection of \src\ at 300\,MHz. This detection limits the size, and thus age, of a remnant due to the requirement that the environment is optically thin to free-free absorption at 300\,MHz.
As such, we consider this highly absorbed scenario unlikely for \src, though still consider it here for 
comparison to past limits on \repeater.

Our burst limits can be compared to those placed for \src\ using other 
telescopes. \citet{tvc+20} place a 3-$\sigma$ persistent 0.3--10\,keV X-ray flux of $5.5\times10^{-14}$\,\fluxcgs\ using 10\,ks of \swift/XRT observations during active periods of \src. For our corresponding limit we use a more stringent 5-$\sigma$ confidence interval. Our 3-$\sigma$ limit, however, would be $4\times10^{-15}$\,\fluxcgs, just over an order of magnitude deeper than the \swift/XRT limit. Using {\em INTEGRAL}/IBIS, \citet{psf+20} place 3-$\sigma$ upper limits on the 28--80 keV gamma-ray flux of $3.4\times10^{-8}$\,\fluxcgs\ for 100-ms-long bursts at anytime during the {\em INTEGRAL} observations. This is very similar to the 10--100\,keV \fermi/GBM limit placed here on gamma-ray emission at the time of radio bursts (translated to a 3-$\sigma$ limit on flux it is $3\times10^{-8}$\,\fluxcgs).

\subsection{Comparison to FRB Models}

We can compare our X-ray and gamma-ray energy limits to the energy emitted by the 2004 giant flare of SGR~1806$-$20, the most energetic event detected from a Galactic magnetar. 
Though most interesting in the context of the magnetar model, this event is
the most luminous transient event yet detected in our Galaxy, so is therefore 
interesting in a model-agnostic context as well.
The bright onset of the flare had a spectrum similar to that of our canonical giant flare model, an isotropic gamma-ray luminosity of $\sim10^{47}$\,\lumcgs\ \citep[measured in the $\sim20$\,keV--10\,MeV band;][]{mca+05,pbg+05}, and a duration of $\sim100$\,ms. This gives an emitted
energy in a 10\,keV--1\,MeV band of $\sim10^{46}$\,erg.
{\ps Our gamma-ray extrapolated isotropic energy limit for the giant-flare-like cutoff power-law model in Table \ref{tab:lims}} is still an order of magnitude higher than this energy emitted by SGR~1806$-$20. Further, Galactic magnetar activity includes much fainter events. The giant flares from magnetars SGR~0526$-$66 and SGR~1900+14 had peak luminosities of $10^{44-45}$\,\lumcgs, over $100\times$ lower than the SGR~1806$-$20 giant flare. 
Short X-ray bursts from magnetars span far fainter luminosities \citep[$\sim10^{36}-10^{43}$\,\lumcgs; e.g.,][]{gwk+99,gwk+00,sk11}.

For the synchrotron blast wave model of FRBs, \citet{mms19} and \citet{mms19a} predict
an expected maximum fluence for a gamma-ray flare of $\sim10^{-13} - 10^{-12}$\,\flucgs\ for \src. This is far below the detection threshold of either our extrapolated X-ray limits (which would depend heavily on what the spectrum of the gamma-ray flare would be in the soft X-ray band) or our \fermi\ limits.
The above shows that although the distance to \src\ is low for an FRB, it is still much too distant to probe the energies expected for magnetar-like activity.

{\ps The discovery of a 16.35-day periodicity in the radio burst activity of \src\ \citep{aab+20} has recently led to models in which the source --- still in many models a neutron star --- is in an orbit or precessing.  However, the current models do not clearly predict X-ray or gamma-ray emission that would be detectable using current instruments, given the distance to \src.  For example, \citet{mvz20} describe a situation in which the relativistic wind of a pulsar or magnetar impinges on an orbiting planetary companion, creating an Alfv\'en wing that if viewed downstream could be a source of FRBs.  Given that this scenario does not require powerful flares from the neutron star itself, observable X-ray emission at the distance of \src\ is not expected.  \citet{iz20} present a binary `comb' model in which FRBs are produced when the magnetosphere of a neutron star interacts with the wind of a massive stellar companion, but make no specific predictions for the brightness of high-energy emission.  \citet{lbb20} note that a hyper-active magnetar that is driven by fast ambipolar diffusion in the core is expected to precess freely with a period of hours to weeks.  This could explain the periodicity of observed burst activity, but there is no reason to think that the magnetar flares themselves would be intrinsically brighter or dimmer compared to those we have considered above.}

 Persistent X-ray emission from FRB sources could arise from a pulsar wind nebula (if the FRB source is a rotation or magnetically powered pulsar).  We therefore compare our limit to the X-ray luminosity of the Crab Nebula, $10^{37}$\,\lumcgs. This is three orders-of-magnitude lower than our persistent X-ray luminosity limit of $2\times10^{40}$\,\lumcgs. We can also compare our X-ray luminosity limit to the luminosities of the brightest X-ray sources. It is comparable to the luminosities of low-luminosity active galactic nuclei \citep{tw03}, bright high-mass X-ray binaries \citep{sk17}, and ultraluminous X-ray sources \citep{erm+19}. For all of these sources, their luminosity distributions extend well below our limit so we cannot rule out any such association with the source of \src. However, it shows that future observations of FRBs closer than \src\ have the potential to make a detection if any of these objects are associated with the source.
 
{\ps Note that when translating our flux and fluence limits here to limits on luminosity or energy, we assume an isotropic energy release. If the high-energy emission from an FRB source is beamed, the energy emitted would of course be lower as its emitted over a 
narrower solid angle. }
 
{\ps For both prompt and persistent emission, we are only just beginning to probe the brightest of possible counterparts to} repeating FRBs. Even for the closest sources, say at $<100$\,Mpc, ruling out high-energy activity from {\ps most models}, such as that expected from a magnetar, is challenging. It is, however, important to place the most stringent possible limits for 
closer sources, in case there are much more energetic counterparts to repeating FRBs. 

{\ps Late in the preparation of this work, we became aware of the works of \citet{pbp+20} and \citet{tvc+20a} where limits were placed on the high-energy emission of \src\ during its
active phases using \xmm, \swift/XRT and {\em AGILE}. The deep \xmm\ limits placed on the X-ray emission by \citet{pbp+20} at the time of radio bursts using are similar to ours placed here with \cxo. The {\em AGILE} limits probe a higher energy range than we considered here with \fermi/GBM. The persistent X-ray emission limits from \swift\ \citep{tvc+20a} and \xmm\ \citep{pbp+20} are consistent with those we place here.}


\acknowledgments
We thank the Dominion Radio Astrophysical Observatory, operated by the National Research Council Canada, for gracious hospitality and useful expertise. The CHIME/FRB Project is funded by a grant from the Canada Foundation for Innovation 2015 Innovation Fund (Project 33213), as well as by the Provinces of British Columbia and Qu\'ebec, and by the Dunlap Institute for Astronomy and Astrophysics at the University of Toronto. Additional support was provided by the Canadian Institute for Advanced Research (CIFAR), McGill University and the McGill Space Institute via the Trottier Family Foundation, and the University of British Columbia. The Dunlap Institute is funded by an endowment established by the David Dunlap family and the University of Toronto. Research at Perimeter Institute is supported by the Government of Canada through Industry Canada and by the Province of Ontario through the Ministry of Research \& Innovation. The National Radio Astronomy Observatory is a facility of the National Science Foundation operated under cooperative agreement by Associated Universities, Inc.
This work is based on observations with the 100-m telescope of the MPIfR (Max-Planck-Institut f\"{u}r Radioastronomie) at Effelsberg. 
We thank the DSN scheduling team and the Goldstone Deep Space Communication Complex (GDSCC) and  
the Madrid Deep Space Communication Complex (MDSCC) staff for scheduling and carrying out the DSN observations.
A portion of this research was performed at the Jet Propulsion Laboratory, California Institute of Technology and the Caltech campus, under a Research and Technology Development Grant through a contract with the National Aeronautics and Space Administration. U.S. government sponsorship is acknowledged.

\allacks

\bibliographystyle{aasjournal}
\bibliography{frbrefs}

\end{document}